\documentstyle[prd,aps,twocolumn]{revtex}
\input{epsf}

\catcode`\@=11

\def\maketitle2{\par 
\begingroup
\let\cite\@bylinecite
\def\thefootnote{\fnsymbol{footnote}}%
\twocolumn[\@maketitle2\vskip2pc]%
\thispagestyle{plain}\@thanks
\endgroup
\def\thefootnote{\arabic{footnote}}%
\setcounter{footnote}{0}%
\let\maketitle2\relax \let\@maketitle2\relax
\let\@thanks\relax \let\@authoraddress\relax \let\@title\relax
\let\@date\relax \let\thanks\relax \let\@abstract\relax 
\let\@pacs\relax}

\def\abstract#1{\gdef\@abstract{{\par 
\bgroup
\ifdim\prevdepth=-1000pt \prevdepth0pt\fi
\hsize\columnwidth
\dimen0=-\prevdepth \advance\dimen0 by17.5pt \nointerlineskip
\small\vrule width 0pt height\dimen0 \relax}{~~}#1\egroup}}

\def\pacs#1{\gdef\@pacs{{\par 
\bgroup
\hsize\columnwidth \parindent0pt
\ifdim\prevdepth=-1000pt \prevdepth0pt\fi
\dimen0=-\prevdepth \advance\dimen0 by20pt\nointerlineskip
\egroup} PACS numbers:~#1}}

\def\@maketitle2{
\@preprint
\@title
\ifdim\prevdepth=-1000pt \prevdepth0pt\fi
\@authoraddress
\@date
\begin{list}{}{\leftmargin=0.10753\textwidth \rightmargin=\leftmargin
\itemsep=1pc\partopsep=-1pc}
\item\@abstract
\item\@pacs
\end{list}
}

\catcode`\@=12

\begin{document}
\draft
\title{Chaos in Time Dependent Variational Approximations to Quantum
Dynamics}  
\author{Fred Cooper$^1$, John Dawson$^2$, Salman Habib$^1$, and Robert
D. Ryne$^3$} 
\preprint{LA-UR-96-3335}
\address{$^1$Theoretical Division, MS B285, Los Alamos National
Laboratory, Los Alamos, NM 87545}  
\address{$^2$Department of Physics, University of New Hampshire,
Durham, NH 03824}  
\address{$^3$Accelerator Operations and Technology Division, MS H817,
Los Alamos National Laboratory, Los Alamos, NM 87545} 

\date{\today}
 
\abstract{Dynamical chaos has recently been shown to exist in the
Gaussian approximation in quantum mechanics and in the self-consistent
mean field approach to studying the dynamics of quantum fields. In
this study, we first show that any variational approximation to the
dynamics of a quantum system based on the Dirac action principle leads
to a classical Hamiltonian dynamics for the variational
parameters. Since this Hamiltonian is generically nonlinear and
nonintegrable, the dynamics thus generated can be chaotic, in
distinction to the exact quantum evolution. We then restrict attention
to a system of two biquadratically coupled quantum oscillators and
study two variational schemes, the leading order large $N$ (four
canonical variables) and Hartree (six canonical variables)
approximations. The chaos seen in the approximate dynamics is an
artifact of the approximations: this is demonstrated by the fact that
its onset occurs on the same characteristic time scale as the
breakdown of the approximations when compared to numerical solutions
of the time-dependent Schr\"{o}dinger equation.}

\maketitle2
\narrowtext

\section{Introduction}

There are many situations in quantum mechanics and field theory where
one hopes that one dynamical degree of freedom can be considered
``classical'' or ``semiclassical.'' In the dynamics of the early
universe, one usually imagines that gravitational energy can be
transferred to particle production, with the gravitational field being
treated semiclassically, {\em i.e.}, the quantum matter fields evolve
in a background ``classical'' gravitational field, the dynamics of
which is in turn determined from the expectation value of the energy
momentum tensor of the quantum field. Similarly in pair production
from strong electric fields, one attempts to describe the background
electric field ``classically'' and solve for the dynamics of the
quantum degrees of freedom in this background field. The time
dependence of the electric field is governed by a Maxwell equation in
which the right hand side is the average value of the current of the
produced pairs. In this sort of approximation of a quantum system
coupled with a semiclassical degree of freedom such as a coherent
electric or gravitational field, the {\em approximate} dynamics of the
quantum system can become chaotic. This was first described by us, and
termed ``semiquantum chaos'' \cite{semichaos,physicad}. A closely
related result having the same cause is ``semiquantal chaos''
\cite{ps} which occurs in the time-dependent Gaussian approximation
for the dynamics of quantum systems.

What happens in these dynamical approximations is that the time
evolution of the parameters governing the shape of the quantum
mechanical wave function (or functional) becomes sensitive to the
initial conditions. In this paper we will first establish that this
behavior can occur in {\em any} variational approximation to the
quantum dynamics ({\em e.g.}, time dependent Hartree
approximation). We will then focus on exactly the same model system
treated in Refs. \cite{semichaos,physicad}, namely a system of two
coupled oscillators described by the Lagrangian:
\begin{equation}
L = {1\over 2} \dot{A}^2 + {1\over 2} \dot{x}^2 - {1 \over 2} ( m^2 +
e^2 A^2) x^2~.  \label{lagtwo}
\end{equation}
This system of two nonlinearly coupled oscillators arose from studying
the problem of pair production of charged pions in a strong external
electric field \cite{kluger} (quantum fluctuations of the electric
field were ignored). In momentum space, the individual modes of the
pion field displayed chaotic behavior. The two-oscillator problem
results from ignoring all but the $k=0$ mode for the quantum field. In
the Lagrangian (\ref{lagtwo}), the $A$ oscillator represents the
time-dependent electromagnetic field and the $x$ oscillator, the $k=0$
mode of the charged pion field.

Treating the electromagnetic $(A)$ field classically is the standard
first term in a large $N$ expansion \cite{CHKMPA} and is related to
the classic problem treated first by Schwinger \cite{schw} on pair
production from external fields. Because such semiclassical methods
are often used in initial value problems in quantum field theory, we
hope to understand the origin of the chaos by considering a simple
quantum mechanical model. To this must be added the important point
that while accurate numerical solutions to the quantum mechanical
problem are available to test the validity of approximations, such a
luxury is not available in field theory.

The semiclassical calculation is equivalent to a Gaussian variational
approximation to the field theory (see Ref. \cite{diss/deco} for more
details and an explanation of dissipation and decoherence in this
approximation). As we show later, all variational approximations to
quantum dynamics lead to classical Hamiltonian dynamics for the
variational parameters (the Gaussian approximation is a special case
of this general result). Therefore, since the resulting Hamiltonian
dynamics is generically nonlinear, chaos can be present in the
approximate dynamics.

We will consider two variational approximations which are equivalent
to two different assumptions about the fluctuations of the $A$
oscillator.  The first approximation (semiclassical or leading order
large $N$) is the assumption that we can ignore {\em all} quantum
fluctuations of the $A$ oscillator (the quantum mechanical version of
the electromagnetic field). This is equivalent to assuming
\begin{equation}
\langle{A^2 x}\rangle=\langle{A}\rangle^2\langle{x}\rangle.
\end{equation}
The Hartree approximation assumes that we include only Gaussian
fluctuations of {\em both} quantum oscillators. This assumption
implies a factorization of the expectation values as:
\begin{equation}
\langle{A^2 x}\rangle=\langle{A^2}\rangle\langle{x}\rangle.
\end{equation}
Both of these approximations, since they include only Gaussian
fluctuations, are really no different than a particular phase space
ensemble of classical solutions of the equations of motion for the
coupled classical oscillator problem, with a particular initial
condition implementing the uncertainty relation. Consequently, the
same chaos discussed above will also be found in the corresponding
dynamics of the classical Liouville equation when only Gaussian
fluctuations are allowed. This aspect of the Gaussian approximation we
will discuss elsewhere \cite{sh}. 

Our numerical results show that in the Hartree approximation, the
onset of chaos, as a function of parameters of the Hamiltonian, is
marginally delayed as compared to the large $N$ (semiclassical)
approximation. We find that both approximations diverge from the exact
numerical simulation of the Schr\"odinger equation at approximately
the same time. After that time, the Hartree approximation
qualitatively tracks the general features of the exact simulation
better than the large $N$ approximation. By direct comparison with the
exact numerical solution we also find that chaos in the variational
approximations occurs roughly on the same time scale as when these
approximations diverge from the exact numerical solution. As is known
on general grounds \cite{pwm}, expectation values of the full quantum
system (which are the variational parameters of the classical
Hamiltonian dynamics) are insensitive to initial conditions: Our
results are completely consistent with this fact. Our interpretation
of the above results is in accord with that of Sundaram and Milonni
\cite{sumi} who have argued that the chaos seen in the approximate
dynamics is not a fundamental feature of the full quantum dynamics but
simply reflects a breakdown of the approximation scheme. It was
further argued in Ref. \cite{sumi} that the approximations are
unreliable when either the classical equations are already chaotic or
when the approximate dynamics is chaotic. To test the second part of
this statement we explored nonchaotic parameter regimes for the
approximate dynamics (but not too far from the onset of chaos) and
found essentially no improvement in the agreement between the exact
quantum and approximate calculations. Thus the existence of chaos is
insufficient to assess the accuracy of the approximations: apparently
the breakdown time (in terms of natural time scales) is the same
whether chaos is present or not.

It is strong nonlinearity rather than just chaotic dynamics which
leads to the breakdown of the approximations. This is hardly
surprising: the Gaussian approximations are equivalent to truncating a
cumulant expansion at second order. If the exact dynamics is strongly
nonlinear, higher order cumulants are generated and a second order
truncation quickly becomes invalid. For chaotic systems,
Ref. \cite{sumi} provides a simple analytic argument, but the
statement is true more generally.

These results might seem to put very strong constraints on dynamical
mean field approximations in quantum field theory, especially at
strong coupling. However, the field theoretic analog to the above
problem has a very large (formally infinite) number of degrees of
freedom. For example, in the field theoretic case, Eqn. (\ref{lagtwo})
is a radical truncation of the full Lagrangian
\begin{equation}
L= |(\partial_{\mu} -ie A_{\mu}) \phi |^2 - {1 \over 4}
(\partial_{\mu}  A_{\nu}-\partial_{\nu}  A_{\mu})^2 - m^2 \phi^{\dag}
\phi~.                                          \label{lagfield}
\end{equation}
In the leading order large $N$ approximation, $A$ is still treated
classically but it is now coupled to a very large number of
fluctuating degrees of freedom. Before definitive statements regarding
the accuracy of mean field approximations can be made, two issues have
to be clarified. The first has to do with the fact that even though
the individual trajectories of the Fourier modes $\phi_k$ may be
chaotic and far from the exact solution, what really matters is the
summed contribution ({\em i.e.}, the statistics of the distribution of
trajectories) and this may have a much more benign character. The
second point is related to the onset of chaos as the number of degrees
of freedom is varied. The importance of this question was noted by
Ford \cite{JFord} but it has not been studied in any detail in the
literature. Thus, it is still an open question whether chaos in the
mean field approximation in field theory is as serious an obstruction
as suggested by the study of lower dimensional systems.

The rest of the paper is set out as follows. First (Sec. II) we prove
the general result that all variational approximations lead to a
Hamiltonian dynamics for the variational parameters. Then, in Sec. III
we explicitly discuss the Hamiltonian dynamics for the two oscillator
problem, both in the large $N$ and Hartree approximations. In Sec. IV
we briefly describe our numerical approach to the exact solution of
the two coupled oscillator problem. We then compare numerical
simulations of the two variational approximations with the exact
solution of the Schr\"odinger equation. Finally, in Sec. V, we state
our conclusions and discuss the implications of our results.

\section{The Time-Dependent Variational Principle}

The Schr\"odinger equation can be reduced to a system of ordinary
differential equations for some variational parameters by constraining
the wave function to be of a particular form. In this section we show
that any variational calculation of the wave function will lead to a
Hamiltonian dynamics for the variational parameters.

The starting point for a variational calculation is Dirac's action
principle \cite{pamd} which can also be used to derive the
Schr\"odinger equation as shown below. We begin by defining the
action:
\begin{equation}
S = \int_{t_1}^{t_2} ~dt~ \langle\Psi|i{\partial \over \partial t}
- H|\Psi\rangle/\langle\Psi|\Psi\rangle~.    \label{act}
\end{equation}
The time dependent Schr\"odinger equation 
\begin{equation} 
(i {\partial \over \partial t} - H)|\Psi\rangle =0~,
\end{equation}
then follows from the variational principle $\delta S=0$ along with
the boundary conditions $\delta |\Psi(t_1)\rangle = 0;~\delta
|\Psi(t_2)\rangle=0$.

Minimizing the action (\ref{act}) on a restricted variational basis
for the wave function:
\begin{equation}
\Psi \rightarrow \Psi_v (y_i(t)); ~~~~~ \int dt \Psi^*_v
\Psi_v = 1
\end{equation}
leads to an effective action functional defined on the variational
parameters $y_i(t)$: 
\begin{equation} 
\Gamma [y_i(t)] = \int ~dt~\langle\Psi_v| i {\partial \over \partial
t} - H|\Psi_v\rangle~,
\end{equation}
where the wave function is usually given in the coordinate
representation. Extremization of the the effective action via $\delta
\Gamma [y_i] = 0$ yields the dynamical equations obeyed by the
variational parameters.

In order to show that any variational solution leads to a symplectic
Hamiltonian dynamics for the variational parameters (the case of
Gaussians was considered in Ref. \cite{rm}), we consider general trial
wave functions which are completely determined by $n$ time-dependent
functions of the form $y_i(t)$, $i=1,\cdots,n$, and written formally
as
\begin{equation}
\Psi(x,t) = \Psi(x;y_i(t))~.
\end{equation}
Here we choose for simplicity a one dimensional Schr\"odinger equation
with arbitrary potential. Note that the entire time dependence of the
wave functions is contained in the variational functions $y_i(t)$. The
Dirac form of the action is then given by
\begin{eqnarray}
\Gamma[y] & = & \int dt \int_{-\infty}^{+\infty} dx
\Psi^*(x;y(t)) \left\{i {\partial \over \partial t} - H
\right\}\Psi(x;y(t)) \nonumber \\
& = & \int dt \> L(y,\dot{y})~,
\label{eq:Dirac}
\end{eqnarray}
with $H$ given by
\begin{equation}
H = - {1 \over 2} {d^2 \over dx^2 } + V(x)~.
\end{equation}
Given the above parametric form of the wave function, $L(y,\dot{y})$ is
{\em always} given by a function of the form,
\begin{equation}
L(y,\dot{y}) = \sum_{i=1}^{n} \pi_i(y) \, \dot{y}_i - h(y)~,
\label{eq:Lform}
\end{equation}
where
\begin{eqnarray}
\pi_i(y) = \int_{-\infty}^{+\infty}dx {i \over 2} &\{&
\Psi^{\ast}(x;y) {\partial\over \partial y_i}\Psi(x;y)    \nonumber\\  
&& - \Psi(x;y) { \partial \over \partial y_i }\Psi^*(x;y)\}  
\end{eqnarray}
and
\begin{equation}
h(y) = \int_{-\infty}^{+\infty}dx \Psi^*(x;y)H\Psi(x;y)~.
\end{equation}

Minimization of the action, Eqn. (\ref{eq:Dirac}), leads to Lagrange's
equations:
\begin{equation}
{d\over dt} {\partial L \over \partial \dot{y}_i } - 
  { \partial L \over \partial y_i } = 0 \>, 
    \quad \hbox{\rm for} \quad i = 1, n  \>.
\end{equation}
The equations of motion for $y_i$ can be found easily using the
specific Lagrangian defined in Eqn. (\ref{eq:Lform}),
\begin{equation}
\sum_{j = 1}^{n} M_{i j}(y) \, \dot{y}_j = 
  { \partial h(y) \over \partial y_i } \>,
\end{equation}
where $M_{i j}(y)$ is an anti-symmetric matrix given by
\begin{equation}
   M_{i j}(y) = { \partial \pi_i \over \partial y_j }
   - { \partial \pi_j \over \partial y_i } 
   = - M_{j i}(y)  \>.
\end{equation}
If the inverse of $M_{i j}$ exists, the equations of motion can be put
in a symplectic form:
\begin{equation}
   \dot{y}_i = \sum_{j = 1}^{N} M_{i j}^{-1}(y) 
  { \partial h(y) \over \partial y_j }  \>.
\end{equation}  
Since $M_{i j}^{-1}$ is also anti-symmetric, $h(y)$ is a conserved
quantity: 
\begin{equation}
   {dh(y) \over dt } = 
   \sum_{i} { \partial h \over \partial y_i } \dot{y}_i= 
   \sum_{i j} { \partial h \over \partial y_i }
   M_{i j}^{-1} { \partial h \over \partial y_j }
   = 0~.
\end{equation}
Following Das \cite{das}, we now introduce Poisson brackets by:
\begin{equation} 
\{ A, B \} = \sum_{ij} {\partial A(y)\over\partial y_i }M_{ij}^{-1}
{\partial B(y) \over \partial y_j }~.
\end{equation}
So, for example,
\begin{equation} 
\{ y_i, y_j \} = M_{ij}^{-1}~.
\label{eq:PByy}
\end{equation}
The equations of motion can now be written in terms of these Poisson
brackets: 
\begin{equation}
\dot{y}_i = \{ y_i,h(y) \} = \sum_{j} M_{ij}^{-1} { \partial h \over
\partial y_j } = \sum_{j} \{ y_i, y_j \} { \partial h \over \partial
y_j }~.            \label{ce}
\end{equation}
The antisymmetry of the Poisson brackets is explicit in their
definition (\ref{eq:PByy}). However, they must also obey Jacobi's
identity:  
\begin{equation}
\{ y_i, \{ y_j, y_k \} \} + \{ y_j, \{ y_k, y_i \} \} + \{ y_k, \{
y_i, y_j \} \} = 0~, 
\label{eq:Jacobi}
\end{equation} 
which is satisfied if $M_{ij}$ obeys Bianchi's identity:
\begin{equation}
{ \partial M_{ij} \over \partial y_k } + { \partial M_{ki}\over
\partial y_j } +  { \partial M_{jk} \over \partial y_i } = 0~.
\label{eq:Bianchi}
\end{equation}  
But Bianchi's identity is always satisfied for $M_{ij}$ of the form  
\begin{equation}
M_{ij} = \partial_i \pi_j - \partial_j \pi_i~.
\end{equation}
Thus our definition of the Poisson brackets satisfies Jacobi's
identity, and the set of classical equations of motion (\ref{ce}) are
symplectic.

\section{Hartree Approximation and the Large $N$ Limit}

We have shown that a time-dependent variational approximation always
leads to a Hamiltonian dynamical system for the variational
parameters. Since such a system is generically nonlinear, there is a
strong likelihood of chaos in the phase space of these Hamiltonian
parameters. In this section we derive two different approximations for
the coupled oscillator problem. The first keeps Gaussian correlations
(Hartree approximation) for both oscillators, while the second (large
$N$ approximation) ignores fluctuations in the $A$ oscillator. The
second approximation has been derived previously from a path integral
approach \cite{physicad} by making $N$ copies of the $x$ oscillator
and then taking the large $N$ limit.

The model Hamiltonian that generalizes the two-oscillator problem to
an $N+1$ oscillator system is  
\begin{equation}
H={1\over 2}p_A^2 + \sum_{i=1}^N {1\over 2}p_i^2 + {1\over 2}(m^2+e^2
A^2)\sum_{i=1}^N x_i^2~,    \label{mham}
\end{equation}
where we have introduced an $N+1$ component oscillator $ x_{\mu};~\mu
= 0,1,\cdots,N$ with $x_0=A$ and the other $N$ oscillators labeled by
the roman indices $i=1,2,\cdots,N$. We show below that at large $N$, a
Gaussian ansatz for the wave function reproduces the exact large $N$
limit of the quantum mechanical system. At $N=1$, the Gaussian
approximation reduces to the well known Hartree approximation.

The operator equations of motion following from the Hamiltonian
(\ref{mham}) are
\begin{eqnarray}
\ddot{x}_i  + ( m^2 + e^2 A^2) x_i & = & 0~,  \label{eq:xdot2q} \\
\ddot{A}+ e^2 \sum_i x_i^2 A & = & 0~. \label{eq:Adot2q}
\end{eqnarray}
Taking expectation values of these two equations we obtain
\begin{eqnarray}
\langle\ddot{x}_i\rangle +  m^2\langle x_i\rangle + e^2\langle A^2
x_i\rangle & = & 0  \label{eq:xdot2} \\ 
\langle\ddot{A}\rangle+ e^2\langle x^2 A\rangle & = & 0
\>. \label{eq:Adot2} 
\end{eqnarray}
It was shown in Ref. \cite{physicad} that in the large $N$ limit,
fluctuations of the $A$ oscillator are suppressed by $1/N$ and the
exact equations (\ref{eq:xdot2}) and (\ref{eq:Adot2}) are approximated
by
\begin{eqnarray} 
\langle\ddot{x}_i\rangle +  m^2\langle x_i\rangle + e^2\langle
A\rangle^2\langle x_i\rangle & = & 0~,  \label{eq:xdot22} \\
\langle\ddot{A}\rangle + e^2\langle x^2\rangle\langle A\rangle & = & 0
\>. \label{eq:Adot22} 
\end{eqnarray} 
The semiclassical field $\langle A\rangle$ now has a time-dependent
mass given by the quantum expectation value $\langle x^2\rangle$. The
quantum oscillator $x_i$ has a mass with time dependence controlled by
$\langle A\rangle$. (This system was discussed in detail in
Refs. \cite{semichaos,physicad}.) It is also perfectly clear that the
large $N$ limit is equivalent to treating the $A$ oscillator
classically ({\em i.e.}, ignoring the quantum fluctuations about the
mean value of $A$).

The equations governing $\langle A\rangle$ and $\langle x^2\rangle=G$
when $\langle x\rangle =0$ were shown to be derivable \cite{semichaos}
from the effective classical Hamiltonian:
\begin{equation}
H_{eff}= {1\over 2} p_A^2 + 2\hbar\Pi_G^2 G + {\hbar\over 8G} +
{\hbar\over 2}\left(m^2 + e^2A^2\right)G.    \label{Heff}
\end{equation} 
We will show below that using a Gaussian trial wave function in
Dirac's variational principle and taking the large $N$ limit will lead
to the same effective Hamiltonian (\ref{Heff}) for the evolution of
the expectation values. However, if instead of taking the large $N$
limit, we set $N=1$, and a trial wave function which is a product of
Gaussians in $A$ and $x$, then the equations for the expectation
values become:
\begin{eqnarray}
\langle\ddot{x}_i\rangle +  m^2\langle x_i\rangle + e^2\langle
A^2\rangle\langle x_i\rangle & = & 0~,  \label{eq:xdot2h} \\
\langle\ddot{A}\rangle + e^2\langle x^2\rangle \langle A\rangle & = &
0~. \label{eq:Adot2h} 
\end{eqnarray}
Here $\langle A^2\rangle = \langle A\rangle^2 + D$, and $D$ is the
Gaussian quantum fluctuation of the $A$ oscillator (which also is the
width of the $A$ wave function). In this case we will also get an
effective Hamiltonian description of the dynamics, but with two more
parameters, $D$ and $\Pi_D$. We will compare these two approximate
Hamiltonian dynamics with the numerical simulation of the exact
dynamics.

Our choice for the trial wave function is 
\begin{eqnarray}
\Psi_v(x_{\mu})&=&N \exp[- {1 \over \hbar} (x-q(t))_{\mu}
  (x-q(t))_{\nu} ({G^{-1} \over 4} - i \Pi)_{\mu \nu} \nonumber\\
&& +{i \over \hbar}p_{\mu}(t)(x-q(t))_{\mu}] 
\end{eqnarray}
where the normalization constant is given by
\begin{displaymath}
N= \exp\left[-{1 \over 4}{\rm{Tr}}\ln (2 \pi \hbar G)\right]~.
\end{displaymath}
The variational parameters are related to various expectation values
taken with respect to the variational wave function $ \Psi_v$:
\begin{eqnarray}
q_i(t) &=& \langle\Psi_v | x_i| \Psi_v\rangle~, \nonumber\\
p_i(t) &=& -\langle\Psi_v |i\hbar {\partial \over\partial x_i }|
\Psi_v\rangle~, \nonumber\\
G_{ij}(t) + q_i(t)q_j(t)&=& \langle\Psi_v | x_i x_j| \Psi_v\rangle~,
\nonumber\\ 
2q_i(t) p_j(t) + 4 \Pi_{ik}(t)G_{kj}(t) &=& \langle\Psi_v | x_i
p_j+p_j x_i| \Psi_v\rangle~.  
\end{eqnarray}
The equations for these expectation values are obtained by varying the
effective action, or equivalently from Hamilton's equations using the
effective Hamiltonian.

The effective action for the variational parameters $p,q,G,\Pi$ is 
\begin{equation}
\Gamma = \int dt~ \left\{ \sum_{i=1}^N  p_i \dot{q}_i+ p_A \dot{A}
-\hbar {\rm{Tr}} [\dot{\Pi} G] - H_{eff}\right\} 
\end{equation}
where $\rm{Tr}[AB] = A_{\mu \nu} B_{\nu \mu}$ and the effective
Hamiltonian,
\begin{eqnarray}
H_{eff}&=&\langle\Psi_v | H | \Psi_v\rangle \nonumber\\
&=&\sum_{i=1}^N {p_i^2\over 2}+ { p_A^2 \over 2} + \hbar
{\rm{Tr}}\left[{1\over 8}G^{-1}\right]+ 2 \hbar {\rm{Tr}}[\Pi G \Pi]
\nonumber \\  
&+& [{m^2 \over 2}+ {e^2 \over 2}(A^2+G_{00})]\sum_{i=1}^N (q_i^2 +
G_{ii})~.  \label{Heff2}
\end{eqnarray}
This last equation gives the effective Hamiltonian for the dynamics of
the $N+1$ oscillators in the Hartree approximation. For simplicity (as
was done in Ref. \cite{semichaos}), we now specialize to the case
$q(t)=p(t)=0$. In this case $G$ and $\Pi$ are diagonal (in general,
they are also diagonal to leading order in the $1/N$ expansion). Since
we have $N$ replicas of the $x$ oscillator, the diagonal condition
simply means that $G_{ij}(t)=G(t)\delta_{ij}$. Inserting this
condition in (\ref{Heff2}) we find
\begin{eqnarray}
H_{eff}^{(0)}&=&{1\over 2}p_A^2+2\hbar\left(N\Pi_G^2G + \Pi_D^2D\right)
+ {\hbar\over 8}\left({N\over G} + {1\over D}\right) \nonumber\\
&& + {\hbar N\over 2}\left[m^2 + e^2\left(A^2 + \hbar
D\right)\right]G. \label{Heff3} 
\end{eqnarray}   
Setting $N=1$ in (\ref{Heff3}), we find the effective Hamiltonian that
controls the Hartree approximation:
\begin{eqnarray}
H_{H}^{(0)}&=&{1\over 2}p_A^2+2\hbar\left(\Pi_G^2G + \Pi_D^2D\right)
+ {\hbar\over 8}\left({1\over G} + {1\over D}\right) \nonumber\\
&& + {\hbar\over 2}\left[m^2 + e^2\left(A^2 + \hbar D\right)\right]G~,
\label{HHB}
\end{eqnarray} 
where $G=G_{11}$ and $D=G_{00}$.

Next we take the large $N$ limit of (\ref{Heff3}) using the same
scaling argument as in determing the large $N$ limit of the path
integral formulation \cite{physicad}: We let $A \rightarrow  \sqrt{N}
\tilde{A}$ and $p_A \rightarrow  \tilde{p}_A$ (leaving invariant  $eA=
\tilde{e} \tilde{A}$). Dividing the effective Hamiltonian by $N$ and
keeping the leading term, we find that the large $N$ Gaussian
effective Hamiltonian is exactly the same as the effective Hamiltonian
found from the leading order large $N$ action \cite{physicad}. The
rescaled effective Hamiltonian reads
\begin{eqnarray}
\tilde{H}_{eff}^{(0)}&=&H^{(0)}_{eff}/N \nonumber\\
&=&{1\over 2}p_A^2+2\hbar\Pi_G^2G + {\hbar\over 8G}
+ {\hbar\over 2}\left(m^2 + e^2A^2\right)G~, \label{Hrescale}
\end{eqnarray}
which is in complete agreement with (\ref{Heff}). (Tildes denoting the
rescaled variables have been suppressed above.) At $N=1$, the Hartree
approximation has two more variational parameters $D$ and $\Pi_D$
compared to the large $N$ approximation. These are related to the real
and imaginary part of the width of the wave function for the $A$
oscillator and are obviously not incorporated in the large $N$
approximation. Because of the extra degrees of freedom incorporated in
it, one might anticipate Hartree to be the better of the two
approximations.

In the Hartree approximation, the Hamilton's equations for the
expectation values are:
\begin{eqnarray}
\dot{A}&=&p_A~,~~~~~~~~~~\dot{p}_A=-e^2\hbar AG~,\\
\dot{G}&=&4\hbar \Pi_GG~,~~~~\dot{D}=4\hbar\Pi_DD~,\\
\dot{\Pi}_G&=&{\hbar\over 8G^2}-2\hbar\Pi_G^2-{1\over 2}m^2-{1\over
2}e^2\hbar\left(A^2+\hbar D\right)~,\\
\dot{\Pi}_D&=&{\hbar\over 8D^2}-2\hbar\Pi_D^2-{1\over 2}e^2\hbar^2 G~.
\end{eqnarray}
In the leading order large $N$ approximation, $D=0$, and there is no
equation for $\Pi_D$. 

For numerical work it is sometimes convenient to switch to a set of
coordinates where the kinetic terms have the usual canonical
form. Defining $\rho_G^2=G$ and $\rho_D^2=D$, the new Hamiltonian is 
\begin{eqnarray}
H_{H}^{(0)}&=&{1\over 2}p_A^2+{1\over 2}p_G^2+{1\over 2}p_D^2
+ {\hbar\over 8}\left({1\over \rho_G^2} + {1\over \rho_D^2}\right)
\nonumber\\ 
&& + {\hbar\over 2}\left[m^2 + e^2\left(A^2 + \hbar
\rho_D^2\right)\right]\rho_G^2~,   \label{canh}
\end{eqnarray}
with the resulting equations of motion
\begin{eqnarray}
\dot{A}&=&p_A~,~~~~~~~~~\dot{p}_A=-e^2\hbar A\rho_G^2~,\\
\dot{\rho}_G&=&p_G~,~~~~~~~\dot{\rho}_D=p_D~,\\
\dot{p}_G&=&{\hbar\over 4\rho_G^3}-(m^2+e^2\hbar\left(A^2+\hbar
\rho_D^2\right)) ~,\\ 
\dot{p}_D&=&{\hbar\over 4\rho_D^3}-e^2\hbar^2
\rho_D\rho_G^2~.\label{caneom} 
\end{eqnarray}
Again the equations for leading order large $N$ are obtained by
setting $\rho_D=0$ and dropping $p_D$. The advantage of this form is
the ease in writing symplectic integrators and also simplifying the
form of the matrices needed to compute the Lyapunov exponents.

The above equations can now be solved numerically. Chaos (in the sense
of nonzero Lyapunov exponents) exists for large enough values of
$e^2$ and for energy sufficiently above the ground state energy.

\section{Semiquantum (Gaussian) Chaos}
\subsection{Numerical Methods}

In this section we display evidence that both the large $N$ and
Hartree approximations are chaotic for appropriate values of the
energy $E$ and the coupling $e$. (In Ref. \cite{semichaos}, the large
$N$ approximation alone was shown to be chaotic.) The dynamics of test
trajectories in the above approximations was studied using a
fourth-order symplectic integrator. (This integrator was implemented
using the second set of variables defined at the end of the last
section.)  Chaos was characterized quantitatively by measuring the
Lyapunov exponent for different initial conditions using standard
techniques \cite{wolf}.

In order to check whether the chaos seen in the approximation is of
some relevance to the full quantum problem, a numerical solution of
the corresponding time-dependent Schr\"{o}dinger equation is
required. This was accomplished by using second and fourth-order
unitary, split-operator, spectral solvers that we have recently
implemented on a large parallel computer \cite{hr}. By using large
grids (up to $4096\times 4096$) sufficient resolution is achieved to
accurately evolve the wave function over times long enough to
meaningfully compare with results from the variational approximations.

The phase space of the large $N$ and Hartree approximations was
characterized using Poincar\'{e} sections. At relatively low energies
and modest values of the coupling constant $e$, both the
approximations led to integrable dynamics. However, increasing either
the energy or the coupling constant quickly led to
nonintegrability. While not carrying out an exhaustive analysis, we
did isolate parametric regions where the chaos was relatively soft
(the area of stochastic orbits was small compared to the area occupied
by regular orbits) and regions where the dynamics was predominantly
chaotic. We also ran a large set of initial conditions to sample the
regions in coupling constant/energy space where the approximations
were regular. This was accomplished by implementing a parallel code to
compute the Lyapunov exponents for a large set of independent
trajectories.

\subsection{Numerical Results}

There are two separate but related questions concerning the
variational approximations. The first question relates to how well
they track the exact numerical calculations. We find that the
approximations break away from the exact calculation on a short time
scale independent of whether they are chaotic or not. However,
qualitative agreement with the numerical results is much better in the
nonchaotic case. The second question refers to the stability of the
approximate solutions as well as the exact solution. In the chaotic
regime of the approximations, the approximate evolution is sensitively
dependent on initial conditions whereas the exact evolution is
not. After a finite time, two approximate evolutions starting from
almost identical initial conditions become completely different in the
chaotic case and no longer bear any phase relationship amongst
themselves or to the exact solution. This is in contrast with the
behavior in the integrable case.

The addition of variational parameters has two effects: it
qualitatively improves the long time behavior in both the regular and
chaotic regimes even though the break time from the exact behavior is
not affected. Secondly, there is some evidence that the onset of chaos
is delayed as more parameters are added and that the value of the
maximum Lyapunov exponent is also decreased. However, an exhaustive
study would require a systematic method of adding variational
parameters for the trial wave functions and this we leave to the
future.

\vspace{1cm}
\epsfxsize=6cm
\epsfysize=4.5cm
\centerline{\epsfbox{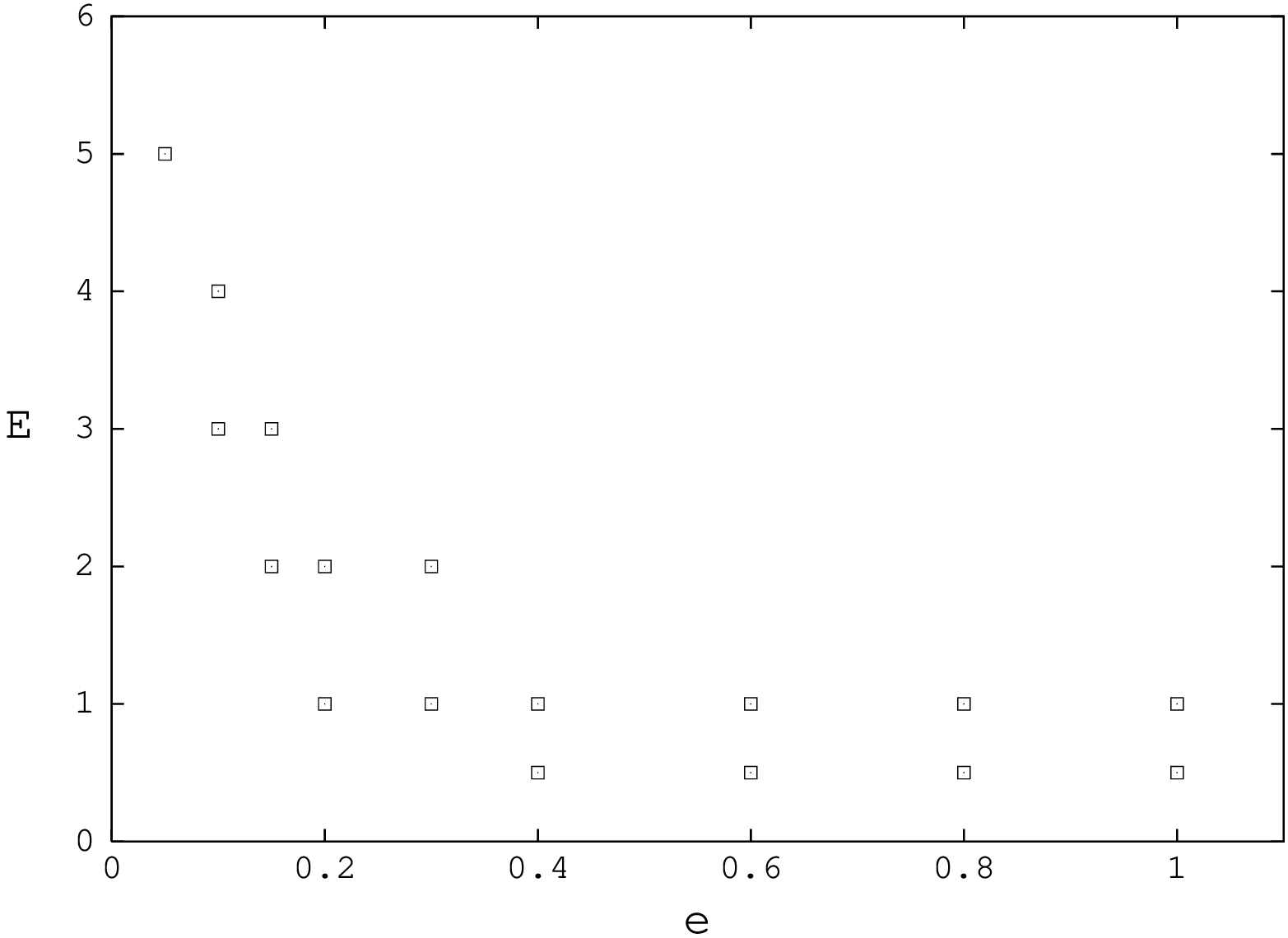}}
\vspace{.35cm} 
{FIG. 1 {\small{Domain of integrability for the large $N$
approximation in energy and coupling constant space. The phase space
was sampled with ten initial conditions at each $(e,E)$ point(and
trajectories asymptoting to positive Lyapunov exponents were searched
for. At fixed $e$, the region above any point denoted by the top
square in the figure corresponds to chaotic dynamics, {\em i.e.} in
the set of trajectories sampled there was at least one with
asymptotically positive Lyapunov exponent. The region below the bottom
square corresponds to integrable dynamics.}}}\\

In Fig. 1 the approximate region of regularity for the large $N$
approximation is displayed. Each $(e,E)$ point was sampled by ten
trajectories, with the Lyapunov exponent calculated for each. Within
the uncertainties of our sampling scheme the integrable and
nonintegrable regions cannot be sharply distinguished: the top set of
points denotes at least one trajectory having an asymptotically
positive Lyapunov exponent while below the bottom set of points no
such trajectory was ever found. The true boundary should be roughly in
the middle of these two curves. The results for the Hartree
approximation are very similar and slightly above the integrability
curve for large $N$ but the difference is of order the uncertainty
band. Whether there is a general (monotonic) tendency for this to
happen as the number of degrees of freedom is further increased is an
interesting speculation which needs to be explored further.

\vspace{1cm}
\epsfxsize=6cm
\epsfysize=4.5cm
\centerline{\epsfbox{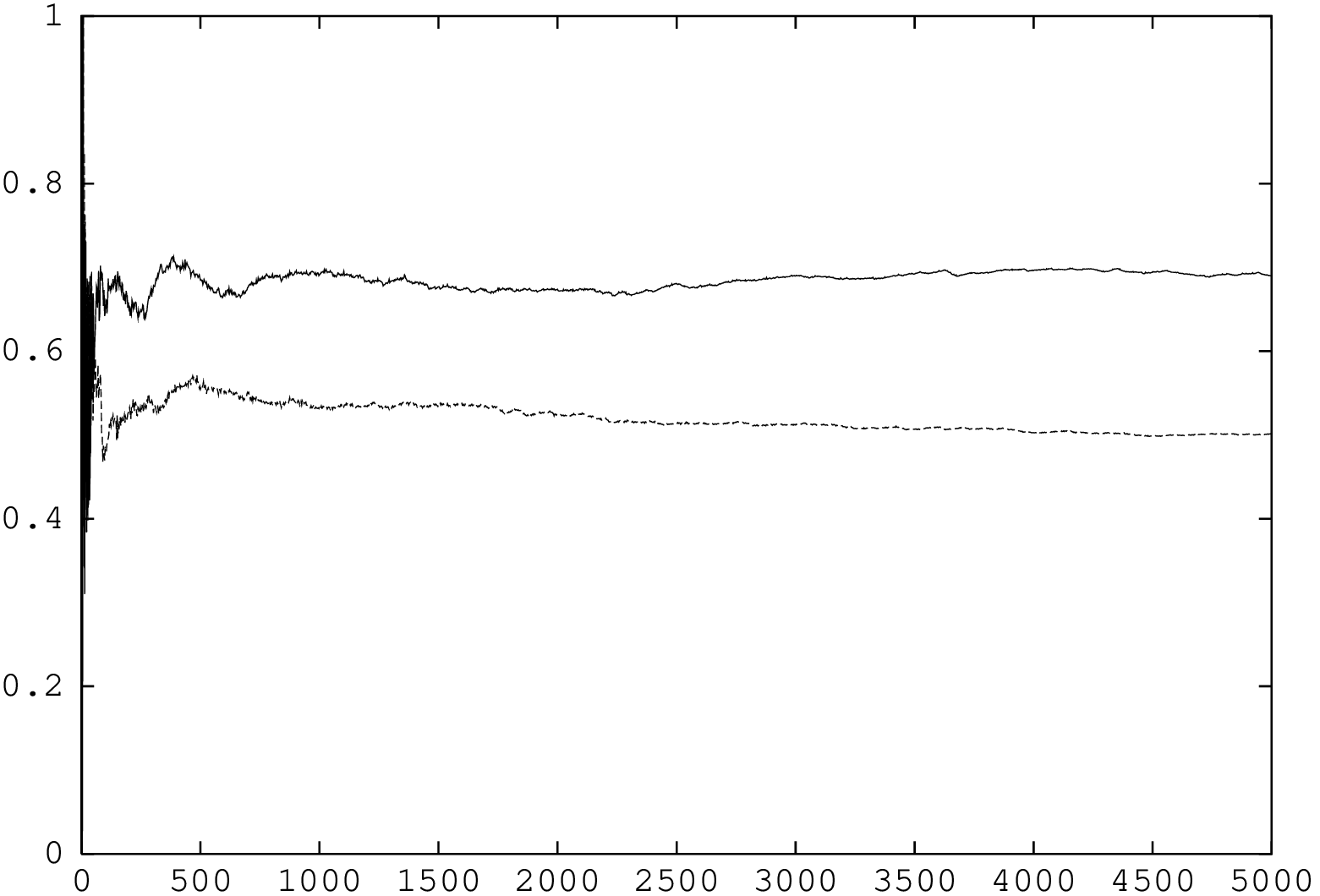}}
\vspace{.35cm} 
{FIG. 2 {\small{A typical computation of the maximal Lyapunov
exponents for the large $N$ (upper curve) and Hartree (lower curve)
approximations. Parameter values for this run were $e=1$ and
$E=5$.}}}\\

The Lyapunov exponents for the two approximations were computed in the
chaotic parameter regime. For all cases we studied the maximal
exponent in the Hartree approximation was less than the corresponding
exponent in the large $N$ approximation. A typical example of these
results is given in Fig. 2.

Poincar\'{e} sections are another way to explore the domains of
integrability for the two approximations. For the ``boundary'' regions
of Fig. 1, the phase space was largely mixed, with stochastic regions
coexisting with regular regions. We checked for random values of the
parameters that the region below this boundary was regular. Above, it
was dominantly chaotic. It was difficult to use Poincar\'{e} sections
for the Hartree approximation because more degrees of freedom means
running much longer to get acceptable statistics. We did run checks
for a few parameter values and found results consistent with Fig. 1
including the fact that chaos occurred at larger values of the
parameters. For example, while the large $N$ approximation had
appreciably chaotic regions at $E=.8$, $e=.7$ the Hartree
approximation was completely integrable for those values of the
parameters. (Note that the energy $E$ is different for the large $N$
and Hartree approximations since $D$ and $\Pi_D$ contribute in the
Hartree approximation, but not in large $N$.) For the parameter
values, $e=1$ and $E=.8$ we show two Poincar\'{e} sections in Figs. 3
and 4 (large $N$) which are typical for values of the parameters near
the boundaries of Fig. 1.

\epsfxsize=6cm
\epsfysize=5cm
\centerline{\epsfbox{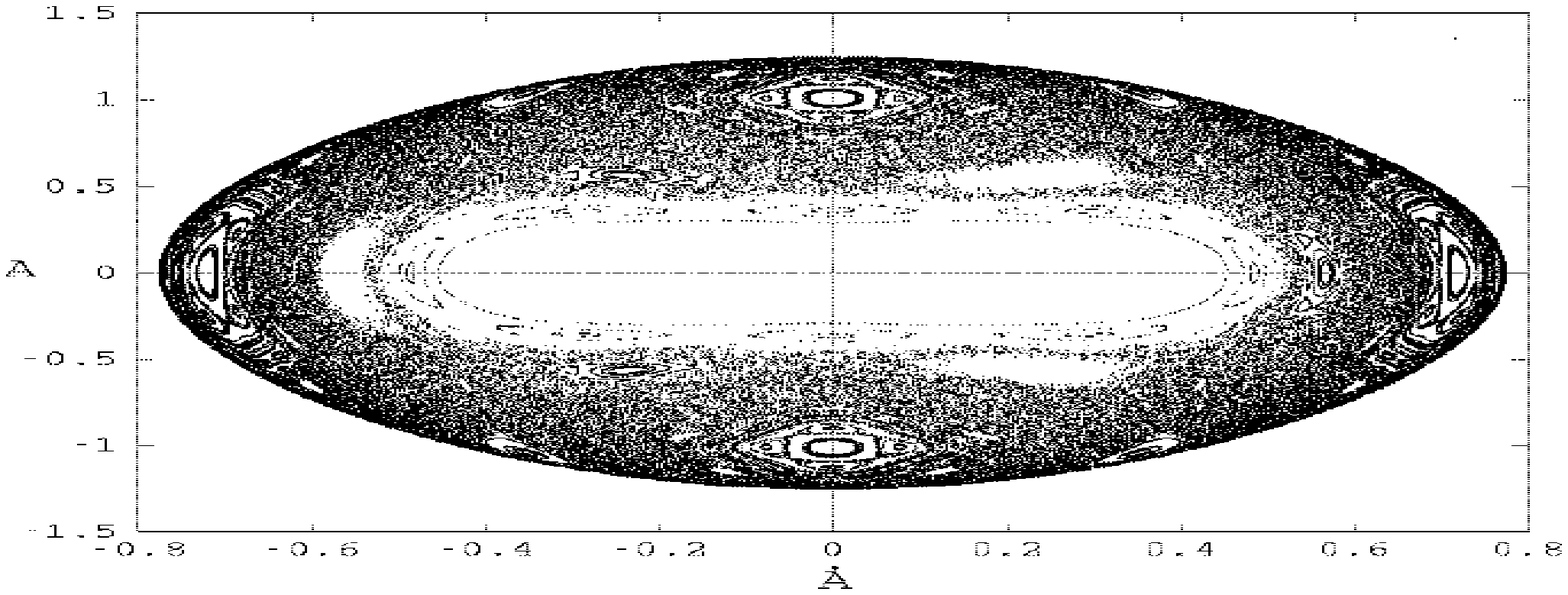}}
\vspace{.35cm} 
{FIG. 3 {\small{A Poincar\'{e} section in the $A, ~\dot{A}$ plane for
$e=1$ and $E=0.8$ for the large $N$ approximation. The phase space was
sampled by 256 different trajectories.}}}\\

\epsfxsize=6cm
\epsfysize=7cm
\centerline{\epsfbox{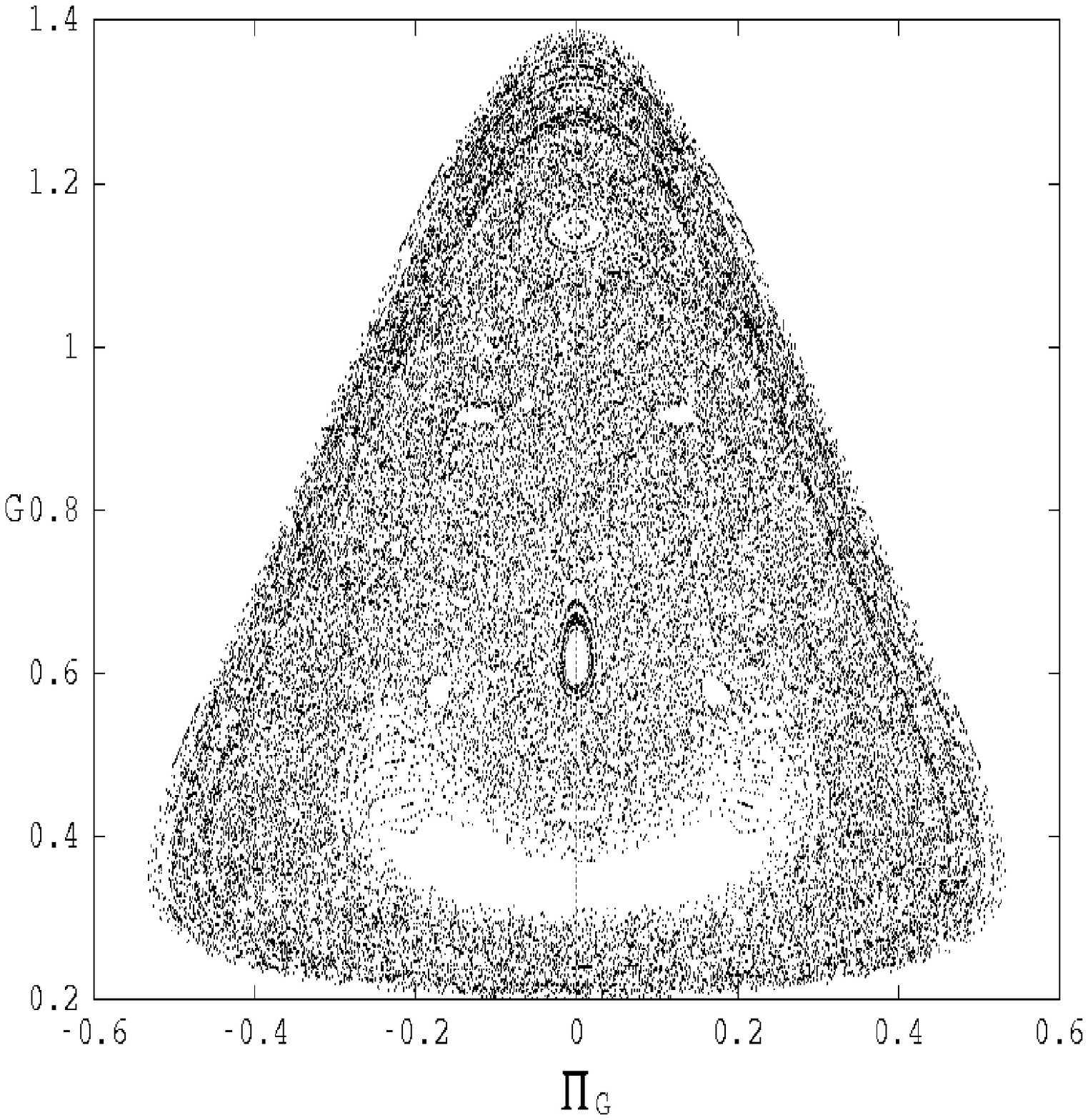}}
{FIG. 4 {\small{A Poincar\'{e} section in the $G, ~\Pi_G$ plane in the
large $N$ approximation for the same set of parameters as Fig. 3.}}}\\

In order to assess the relevance of the chaos seen in the
approximations we have compared the approximate evolutions with exact
numerical solutions of the Schr\"odinger equation with Gaussian
initial data. The exact evolution shows no hint of the sensitivity to
initial conditions exhibited by the approximate dynamics. As
illustrated in Figs. 5, 6, and 7, in both the regular and chaotic
regimes the approximations quickly deviate from the exact results on a
time scale of order unity, this signaling the breakdown of the
Gaussian approximation.

The Lyapunov time sets a maximum time for which the approximations can
agree with the exact quantum dynamics. In fact, consistent with this
statement we observe that the time of breakdown of the approximations
and the Lyapunov time are of the same order. However, this should not
lead one to conclude that the accuracy dramatically improves when the
approximate dynamics is integrable. Indeed, even in integrable
parameter regimes, the breakdown time can remain of order unity
(Fig. 5). Therefore, for coupling constants of order unity, these
approximations tend to be rather poor. This is because significant
non-Gaussian structure forms in the exact wave functions relatively
rapidly.

\vspace{.4cm}
\epsfxsize=6cm
\epsfysize=4.5cm
\centerline{\epsfbox{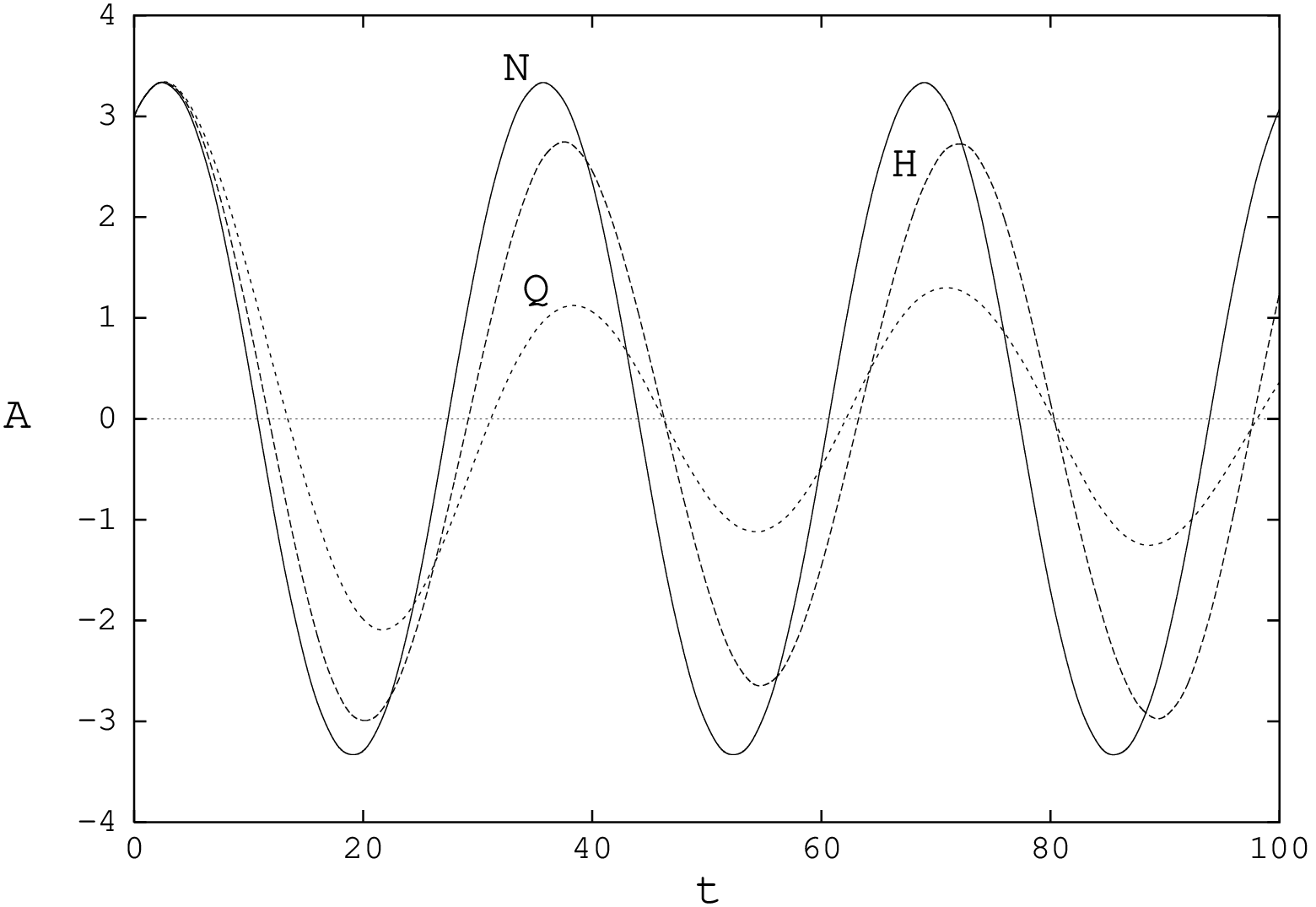}}
\vspace{.35cm} 
{FIG. 5 {\small{Evolution of $\langle A\rangle$ for $e=.3$ and
$E=1$. This is within the parameter range for nonchaotic evolution
within the approximations. Both approximations break away from the
exact evolution at $t\sim 5$ but stay in phase at later times. The
Hartree approximation ($H$) does better in tracking both phase and
amplitude. The curve with the smallest average amplitude corresponds
to the exact quantum evolution ($Q$), and the one with the largest
average amplitude corresponds to the large $N$ expansion ($N$).}}}\\

\vspace{.4cm}
\epsfxsize=6cm
\epsfysize=4.5cm
\centerline{\epsfbox{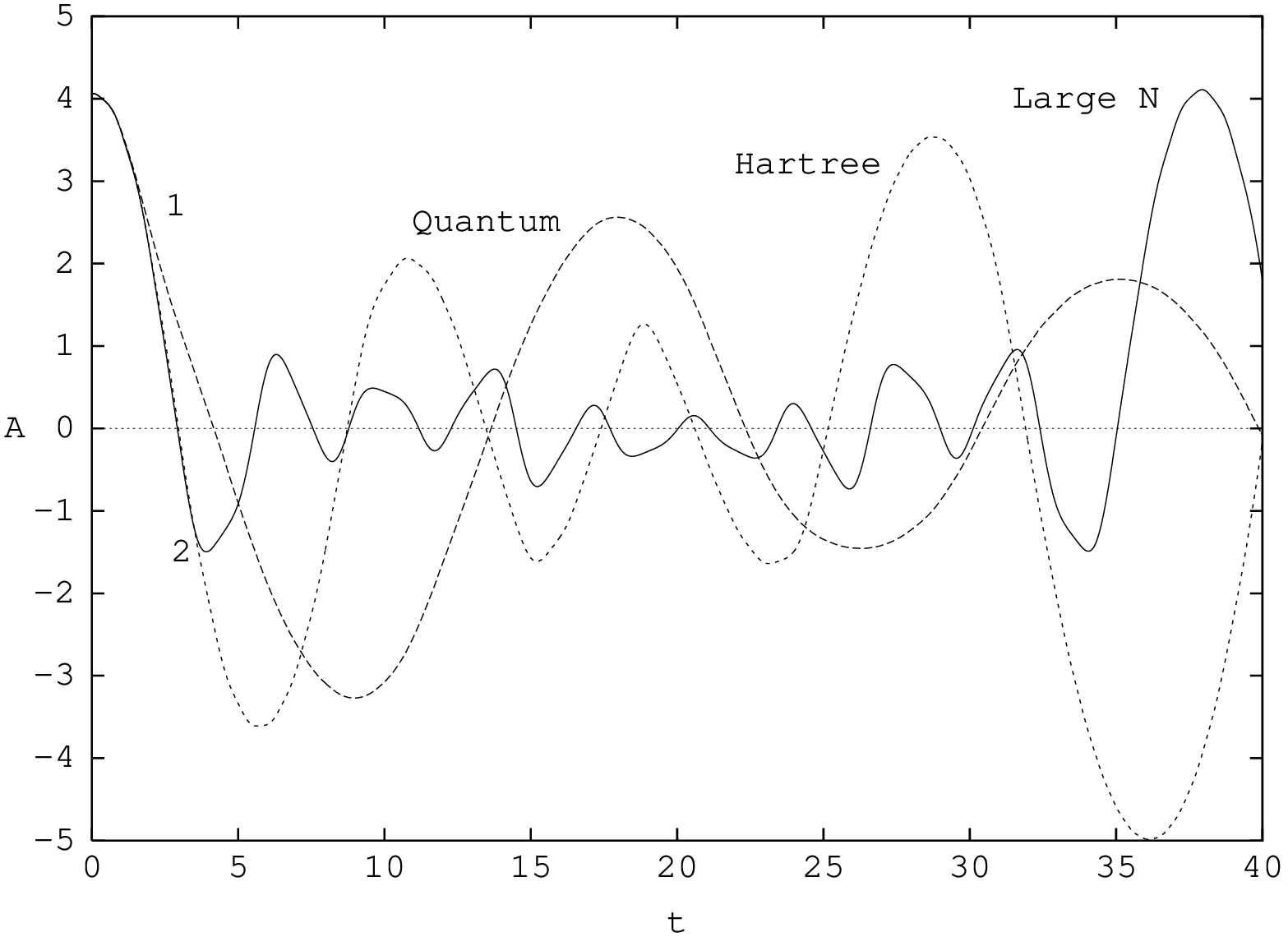}}
\vspace{.35cm} 
{FIG. 6 {\small{Evolution of $\langle A\rangle$ for $e=1$ and
$E=5$. The approximate evolutions are now chaotic. They break away
from the quantum evolution at time $t\sim 2$ (denoted by the point 1
in the figure) and break away from each other at point 2 ($t\sim
4$). In this case the evolutions quickly dephase from each other and
from the quantum evolution.}}}\\

The chaos inherent in the approximations is demonstrated in Figs. 8,
9, and 10, for the evolution of $\langle A\rangle$, $G$, and $D$. In
these figures we show two trajectories for each of the approximations,
one corresponding to an initial $G=.5$ and the other to $G=.5001$ (all
other parameters held fixed). The deviations of these two curves are
consistent with the calculated Lyapunov exponent (which is of order
unity) and an initial deviation of order $10^{-4}$.

\vspace{.4cm}
\epsfxsize=6cm
\epsfysize=4.5cm
\centerline{\epsfbox{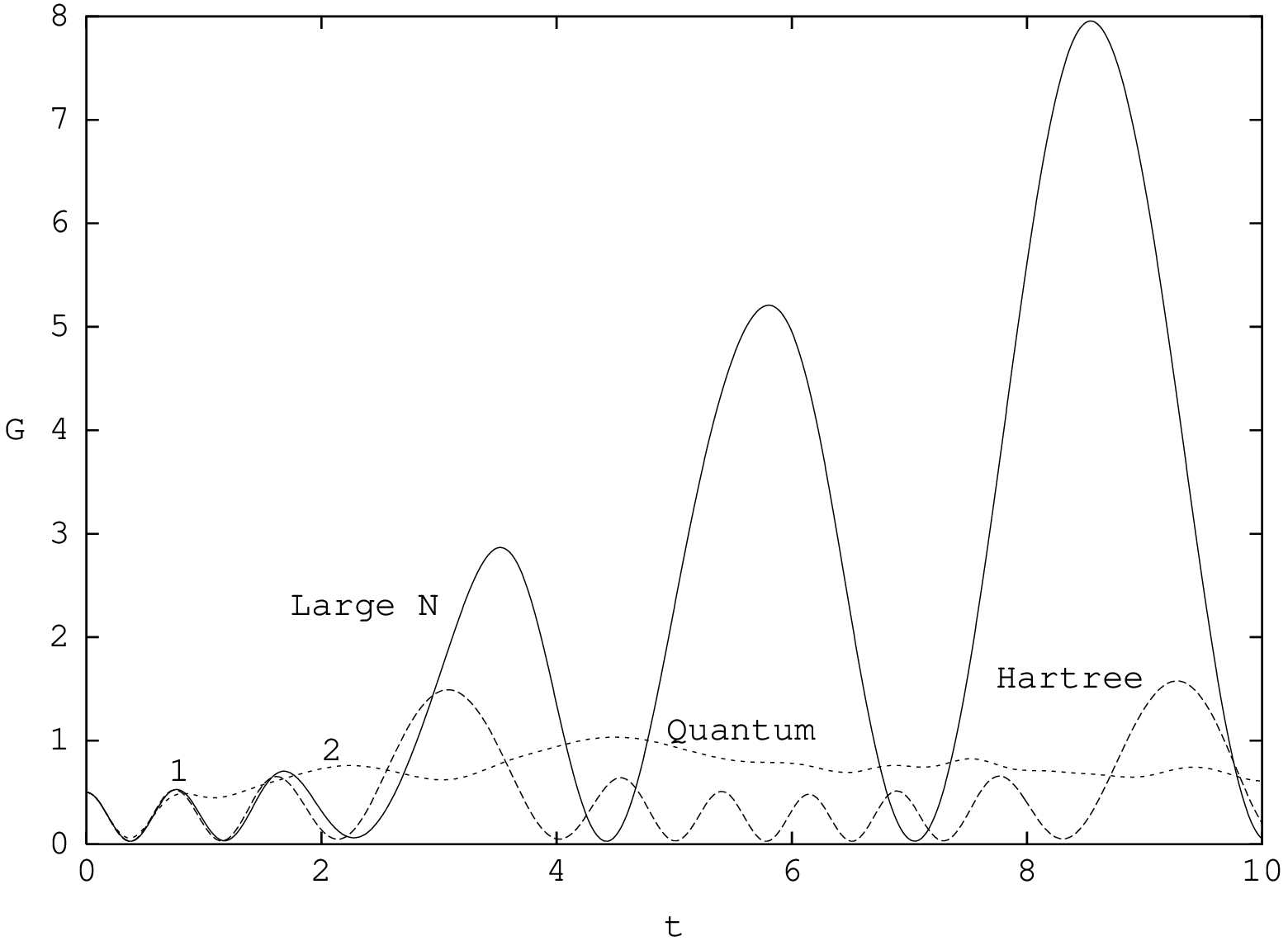}}
\vspace{.35cm} 
{FIG. 7 {\small{Evolution of $G$ for the same parameters as
Fig. 6. The break from the quantum evolution occurs at $t\sim 1$
(denoted by the point 1 in the figure) and the approximations break
away from each other at point 2 ($t\sim 2$). Even though the Hartree
approximation is not correct it does not have the big excursions shown
by the Large $N$ approximation.}}}\\

\vspace{.4cm}
\epsfxsize=6cm
\epsfysize=4.5cm
\centerline{\epsfbox{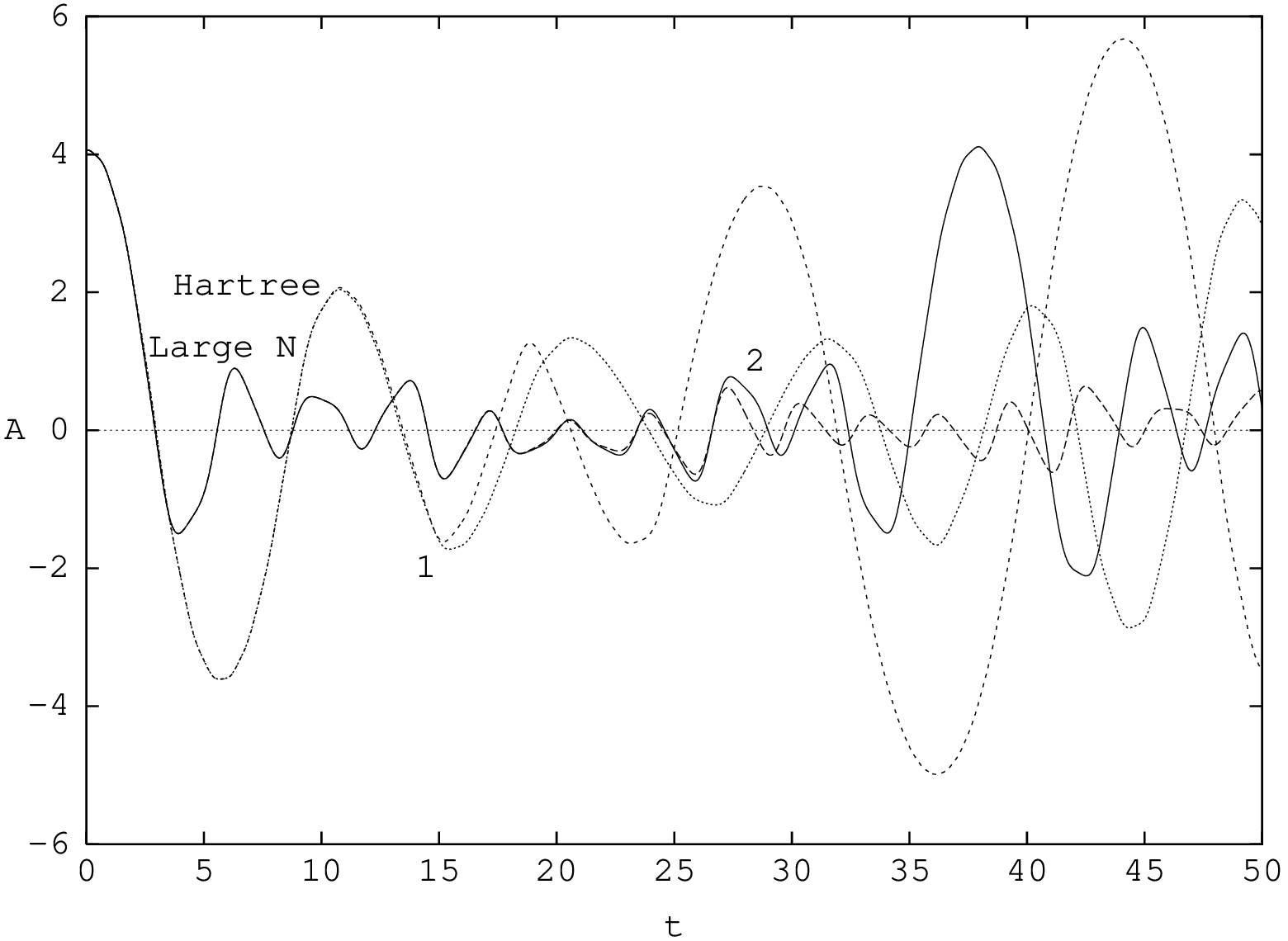}}
\vspace{.35cm} 
{FIG. 8 {\small{Evolution of $\langle A\rangle$ for the same
parameters as Fig. 6. Points 1 and 2 mark the breaking away of two
nearby trajectories in the Hartree and Large N approximations. After
this time, the trajectories rapidly dephase from each other.}}}\\

\vspace{.4cm}
\epsfxsize=6cm
\epsfysize=4.5cm
\centerline{\epsfbox{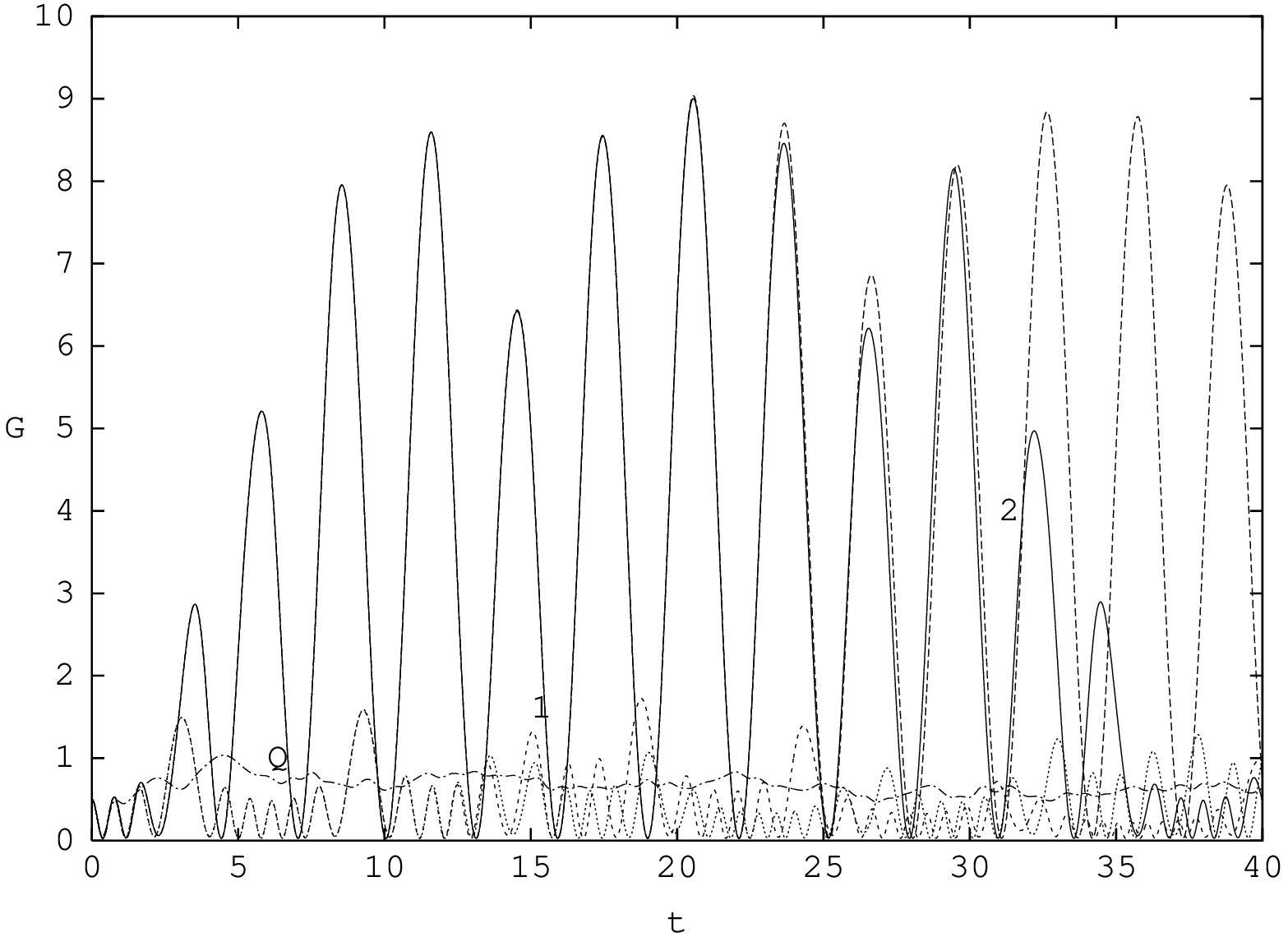}}
\vspace{.35cm} 
{FIG. 9 {\small{Evolution of $G$ for the same parameters as
Fig. 6. The trajectory denoted by $Q$ is the quantum evolution. Points
1 and 2 mark the breaking away of two nearby trajectories in the
Hartree and Large N approximations.}}}\\

\vspace{.4cm}
\epsfxsize=6cm
\epsfysize=4.5cm
\centerline{\epsfbox{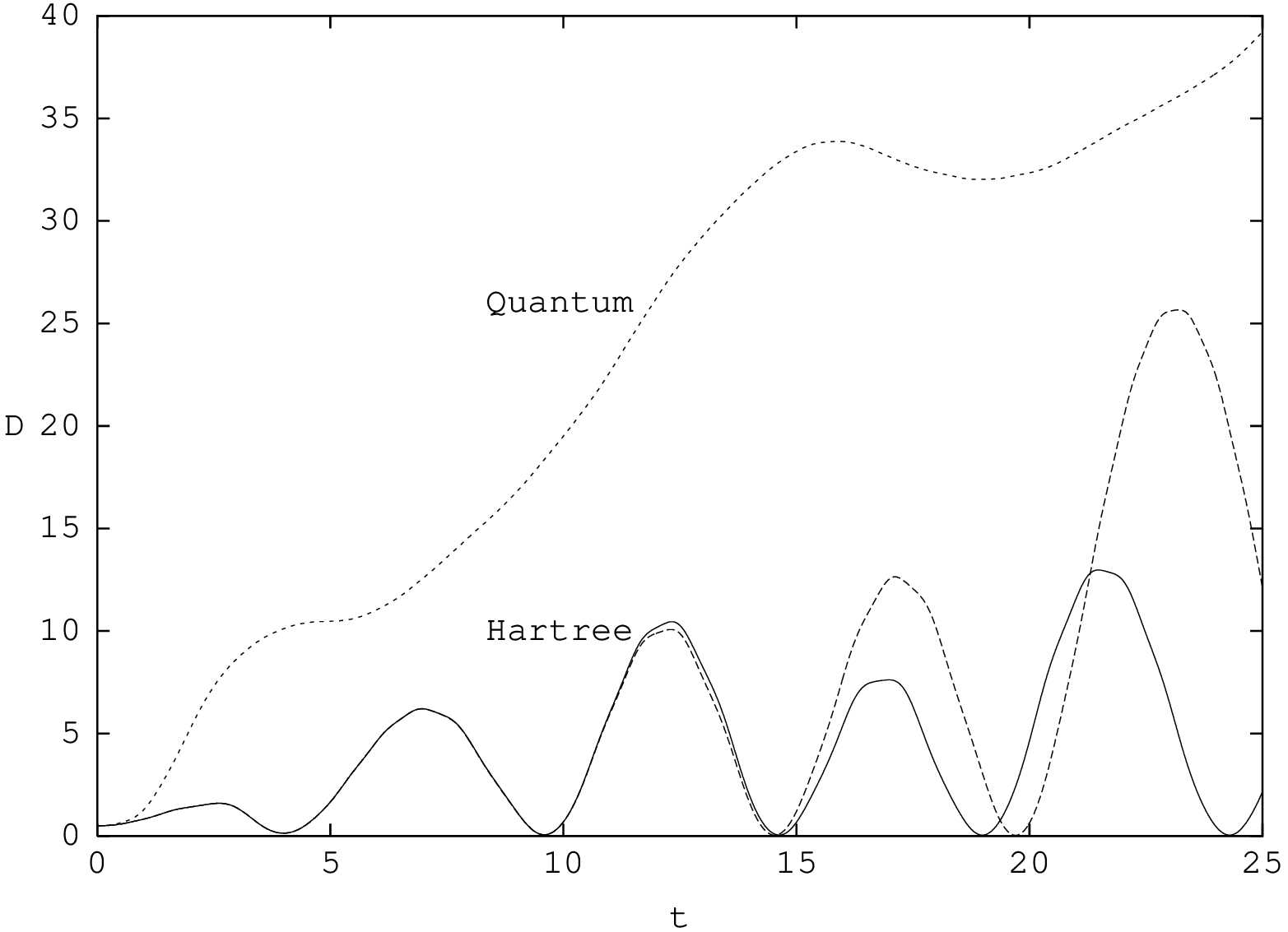}}
\vspace{.35cm} 
{FIG. 10 {\small{Evolution of $D$ for the same parameters as
Fig. 6. The quantum evolution and the Hartree approximation deviate
from each other at $t\sim 1$ and the two nearby trajectories of the
Hartree approximation break from each other at $t\sim 15$.}}}\\

The approximations discussed here break down whenever there is
significant non-Guassian structure in the actual wave function. As
long as the coupling is of order unity this happens relatively
rapidly. Examples of the numerically evaluated probability densities
in $A$ are shown in Figs. 11 and 12 for for values of the parameters
which correspond to integrable and nonintegrable evolutions.

\vspace{.4cm}
\epsfxsize=6cm
\epsfysize=4.5cm
\centerline{\epsfbox{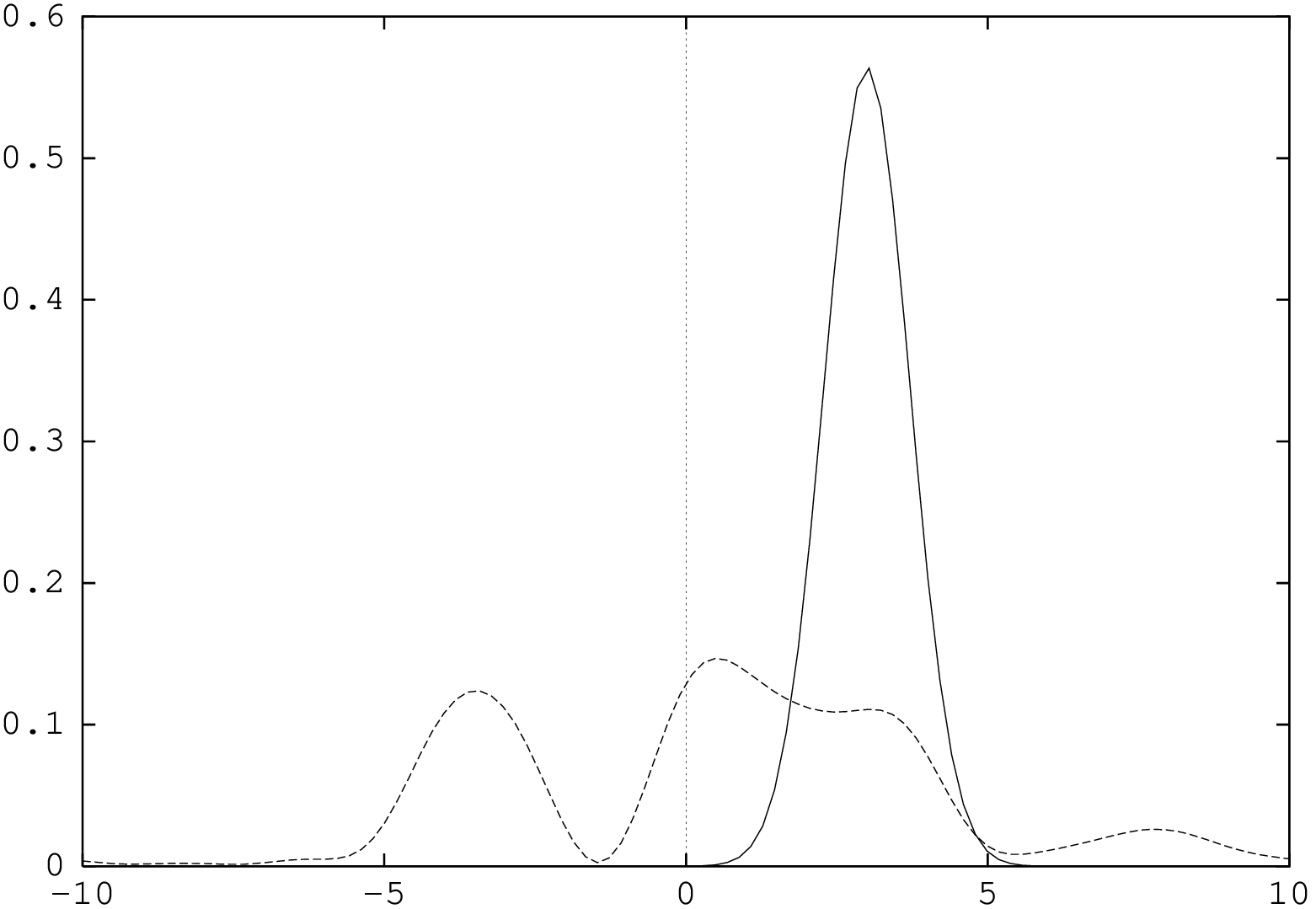}}
\vspace{.35cm} 
{FIG. 11 {\small{The initial and final $(t=100)$ probability densities
for $A$ with $e=.3$ and $E=5$ (integrable case).}}}\\

\vspace{.4cm}
\epsfxsize=6cm
\epsfysize=4.5cm
\centerline{\epsfbox{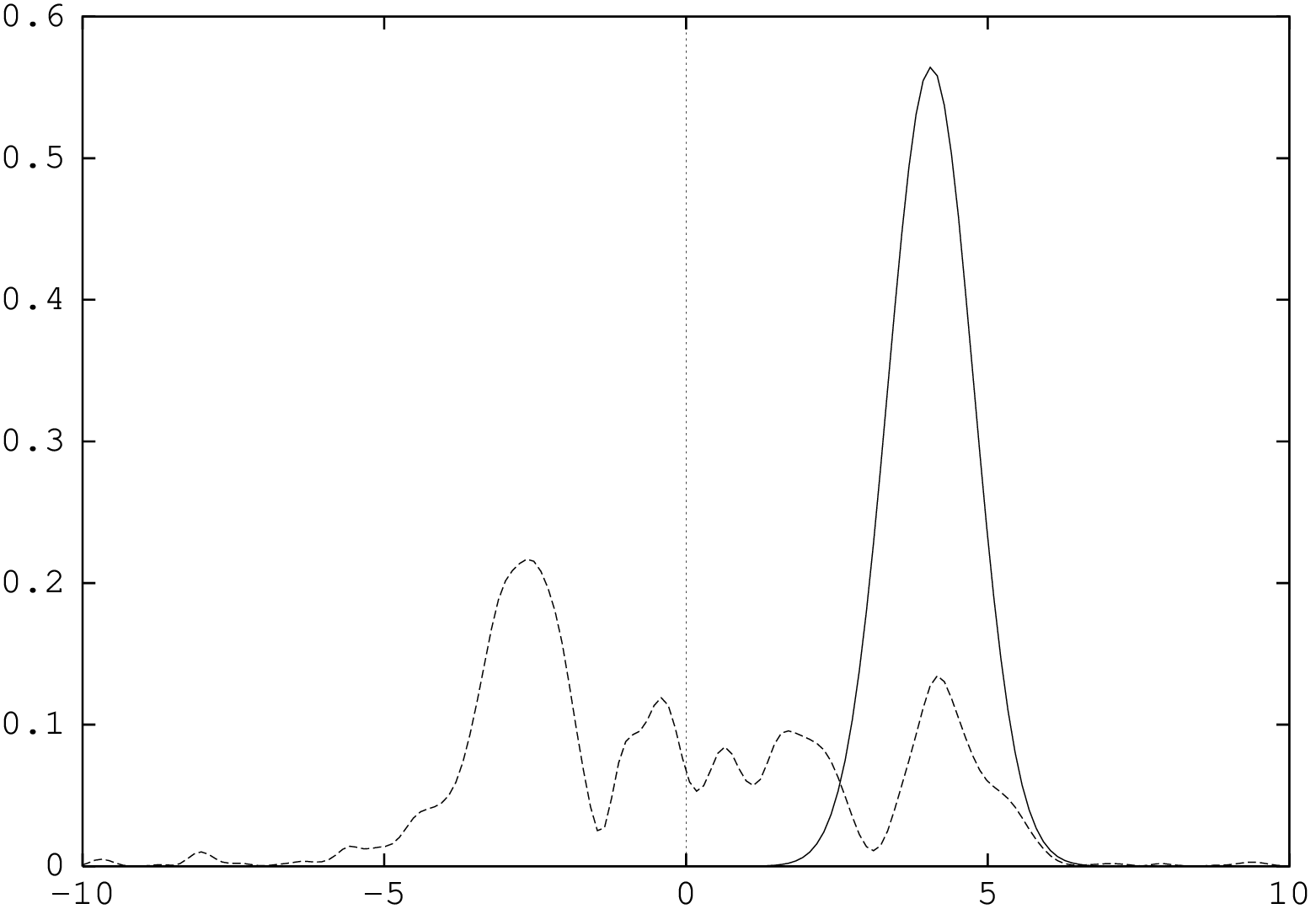}}
\vspace{.35cm} 
{FIG. 12 {\small{Same as Fig. 10 with $e=1$, $E=5$ (nonintegrable
case). In this case the final time, $t=40$.}}}\\ 

\section{Conclusions}

The central results of this investigation may be encapsulated
succinctly: all time-dependent variational approximations based on a
Dirac approach are Hamiltonian and generically nonlinear. Therefore
all such approximations can be chaotic. Since exponential divergence
of expectation values in time is ruled out in full quantum mechanics,
the Lyapunov time associated with the approximate evolution sets a
time scale beyond which the approximation breaks down. We have
investigated this last point in two particular examples (large $N$ and
Hartree for a two-dimensional potential), find both to be chaotic, and
by comparison against numerically obtained solutions, show explicitly
that the approximations break down on the Lyapunov time scale. We also
show that even in nonchaotic regimes, the approximations break down
very quickly. Thus the mere absence of chaos is not an indicator of
the accuracy of these approximations.

We would also like to point out that suggestions have been made in the
literature that semiquantum chaos may in fact be a real effect ({\em
e.g.}, Ref. \cite{ps} and rather more strongly in
Ref. \cite{blel}). However, these claims were not backed up by careful
comparisons with exact calculations. The detailed results reported
here, along with the fact that Gaussian approximations are dynamically
completely classical \cite{sh}, imply exactly the opposite conclusion
(in substantial agreement with the arguments of Ref. \cite{sumi}).

The fact that in the chaotic regime, the approximation signals its own
breakdown has an interesting physical consequence: if the $1/N$
approximation is in fact sensible then a breakdown at leading order
must imply that the next-to-leading terms are becoming large on the
same time scale. Since, in field theory the leading order
approximation does not incorporate collisions, what this implies is
that the collisional time scale can be estimated from the breakdown of
the leading order result itself, without actually having to compute
the next-to-leading order contribution. Given the complexity of
higher-order calculations this feature may be extremely useful. This,
and other aspects of the field theoretic problem are now under
investigation.

One way to of incorporating higher order correlation functions in
dynamical approximations is to consider trial wave functions of the
form Gaussian times polynomials. This can be put in correspondence
with the large $N$ expansion which can be shown to lead to the same
structure. An interesting question is whether opening up the
possibility of including higher order correlations in this way will
improve the long time behavior of the variational approach.

\section{Acknowledgments}

The authors acknowledge helpful conversations with Peter Milonni, Emil
Mottola, Arjendu Pattanayak, Bala Sundaram, and George Zaslavsky. The
large scale numerical work was performed on the CM5 at the Advanced
Computing Laboratory, Los Alamos National Laboratory.

\end{document}